\documentclass[letter,scriptaddress,twocolumn, prl,showkeys,showpacs,superscriptaddress]{revtex4-1}

	\usepackage{amsmath}%,amssymb} 
	\usepackage{makeidx}
	\usepackage{amsfonts}
	\usepackage{graphicx, xcolor}  % needed for figures
	\usepackage{dcolumn}   % needed for some tables
	\usepackage{bm}        % for math
	\usepackage{amssymb}   % for math
	\usepackage[ansinew]{inputenc}
	\usepackage[usenames,dvipsnames]{pstricks}
	\usepackage{subfigure}
	\usepackage{epstopdf}
	\usepackage{pst-grad} % For gradients
	\usepackage{pst-plot} % For axes
	\usepackage[colorlinks,hyperindex]{hyperref}
	\hypersetup
	{
		colorlinks,%
		citecolor=black,%
		linkcolor=black,%
		urlcolor=black,%
	}
	\usepackage{auto-pst-pdf}
	
\newcommand{\be}{\begin{equation}}
\newcommand{\ee}{\end{equation}}
\newcommand{\ba}{\begin{eqnarray}}
\newcommand{\ea}{\end{eqnarray}}
\newcommand{\bml}{\begin{multline}}
\newcommand{\eml}{\end{multline}}

\newcommand{\rd}{{\rm d}}

\begin{document}

\title{A model for the erosion onset of a granular bed sheared by a viscous fluid}
\author{Le Yan\thanks{ly452@nyu.edu}}
\affiliation{Center for Soft Matter Research, Department of Physics, New York University, \\4 Washington Place, New York, 10003, NY}
\author{Antoine Barizien\thanks{marcobaityjesi@gmail.com}}
\affiliation{\'Ecole Polytechnique Universit\'e, Paris-Saclay, France}
\author{Matthieu Wyart\thanks{mw135@nyu.edu}}
\affiliation{Center for Soft Matter Research, Department of Physics, New York University, \\4 Washington Place, New York, 10003, NY}

\date{\today}

\begin{abstract}
We study theoretically the erosion threshold of a granular bed forced by a viscous fluid. We first introduce a novel model
of  interacting particles driven on a rough substrate. It predicts a continuous transition at some threshold forcing $\theta_c$,
beyond which the particle current grows linearly $J\sim \theta-\theta_c$, in agreement with experiments. The stationary state is reached after a transient time $t_{\rm conv}$ which diverges near the transition as $t_{\rm conv}\sim |\theta-\theta_c|^{-z}$ with $z\approx 2.5$. The model also makes quantitative testable predictions for the drainage pattern: the distribution $P(\sigma)$ of local current is found to be extremely broad with $P(\sigma)\sim J/\sigma$, spatial correlations for the current are negligible in the direction transverse to forcing, but long-range parallel to it. We explain some of these features using a scaling argument and  a mean-field approximation that builds an analogy with $q$-models. We discuss the relationship between our erosion model and models for the depinning transition of vortex lattices in dirty superconductors, where our results may also apply. 

%Under a weak external drive, sediments are eroded on a solid bed when the drive is above a critical value. We propose an experimentally testable model of the erosion. The model shows a linear relation between the erosion rate and the external drive above the critical threshold. The linear relation is nontrivially rooted in the fact that the erosion covers the whole landscape, due to a splitting effect of the interaction among eroding particles in dense particle drains. We emphasize this splitting effect with a theoretic analysis on the probability distribution in the local erosion flux.
\end{abstract}

\maketitle
Erosion shapes Earth's landscape, and occurs when a fluid  exerts a sufficient shear stress on a sedimented layer. It is controlled by the dimensionless Shields number $\theta\equiv\Sigma/(\rho_p-\rho)gd$, where $d$ and $\rho_p$ are the particle diameter  and density, and $\rho$ and $\Sigma$ are the fluid density and the shear stress.  Sustained sediment transport can take place above some critical value $\theta_c$ \cite{Shields36,White70,Lobkovsky08}, in the vicinity of which motion is localized on a thin layer of  order of the particle size, while deeper particles are  static or very slowly creeping \cite{Charru04,Aussillous13,Houssais15}. This situation is relevant in gravel rivers, where erosion occurs until the fluid stress approaches threshold \cite{Parker07}. In that case, predicting the flux $J$ of particles as a function of $\theta$ is difficult, both for turbulent and laminar flows \cite{Bagnold66,Charru04}. We focus  on the latter, where  experiments show that: (i)  in a stationary state, $J\propto (\theta-\theta_c)^\beta$ with $\beta\approx 1$ \cite{Charru04,Ouriemi09,Lajeunesse10,Houssais15}, although other exponents are sometimes reported \cite{Lobkovsky08}, (ii) transient effects occur on a time scale that appears to diverge as  $\theta\rightarrow\theta_c$ \cite{Charru04,Houssais15} and (iii) as $\theta\rightarrow\theta_c$ the number of moving particles vanishes, but not their characteristic speed \cite{Charru04,Lajeunesse10}.

Although a continuous description of erosion appears successful for $\theta\gg\theta_c$ \cite{Leighton86,Ouriemi09,Aussillous13}, it should not apply for $\theta\rightarrow\theta_c$. In the latter regime, an erosion/deposition model was proposed in \cite{Charru04}, where one assumes that a $\theta$-dependent fraction of initially mobile particles evolve over a frozen static background, which contain  holes. In this view, $\theta_c$ occurs when the number of holes matches the number of initially moving particles. This phenomenological model, which assumes no interactions between mobile particles, captures (i,ii,iii) qualitatively well. This success is  surprising:  due to the frozen background, one expects mobile particles to take favored paths and to eventually clump together into "rivers", thus avoiding most of the holes. Models including this effect as well as particle interactions \cite{Watson96, Watson97} have been introduced in the context of the depinning transition of vortex lattice in dirty superconductors. They lead to a sharp transition for the flux at some finite forcing, but $\beta\approx 1.5$. Moreover, there are currently no predictions for the spatial organization of the flux near threshold, although this property is  indicative of the underlying physics, and could be accessed experimentally.

In this letter we introduce a model of interacting particles  forced along one direction on a disordered substrate. Particle interactions based on mechanical considerations are incorporated. Such model recovers (i,ii,iii) with $\beta=1$ and an equilibration time $t_{\rm conv}\sim |\theta-\theta_c|^{- 2.5}$. In addition, we find that (a) the spatial distribution of local flux $\sigma$ is extremely broad, 
and follows $P(\sigma)\sim 1/\sigma$ and (b) spatial correlations of flux are short-range and very small in the lateral direction, but are power-law in the mean flow direction. We derive $\beta=1$ and explain why $P(\sigma)$ is broad using a mean-field description of our model, leading to  an analogy with $q$-models \cite{Liu95,Coppersmith96} used to study force propagation in granular packings.

{\it Model:} we consider a density $n$ of particles  on a frozen background. $n$ should be chosen to be of order one, but its exact value does not affect our conclusions. The background is modeled via a square lattice, whose diagonal indicates the direction of  forcing, referred to as ``downhill". The lattice is bi-periodic, of dimension $L\times W$, where $L$ is the length in downhill direction and $W$ the transverse width.  Each node $i$ of the lattice is ascribed a height $h_i\in [0,1]$, chosen randomly with a uniform distribution. Lattice bonds $i\to j$ are directed in the downhill direction, and characterized by an inclination $\theta_{i\to j} = h_i-h_j$. We denote by $\theta$ the amplitude of the forcing. For an isolated particle on site $i$, motion will occur along the steepest of the two outlets (downhill bonds) \cite{Rinaldo14}, if it satisfies $\theta+\theta_{i\to j}>0$. Otherwise, the particle is trapped.

\begin{figure}[!ht]
\centering
\includegraphics[width=1.0\columnwidth]{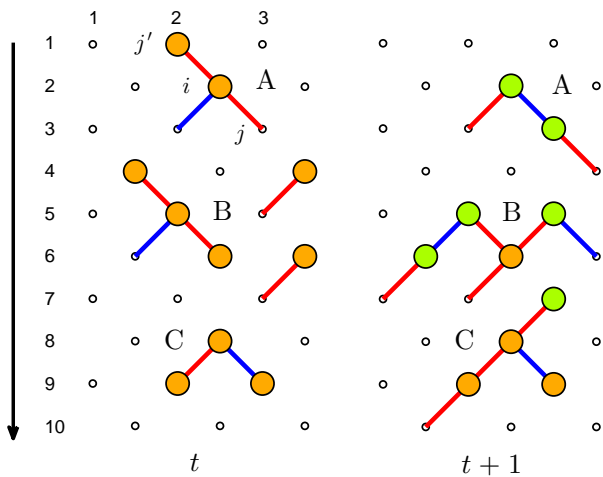}
\caption{\small{Illustration of the model. Small circles indicate lattice sites, particles are represented by discs in yellow, or green if motion occurred between $t$ (left) and $t+1$ (right). The black arrow is in the downhill direction. Solid lines indicate outlet with positive forces. If a particle has two outlets with positive forces, the larger (smaller) one is colored in red (blue). %The red ones show the primary directions and the blue ones are the secondary directions if exist. Particles move according to rules i-iv at each time step. For the specific configuration on the left at time $t$, the particles moved are shown in green at time $t+1$ on the right.
}}\label{model}
\end{figure}

However, if particles are adjacent, interaction takes place. First, particles cannot overlap, so they will only move toward unoccupied sites.  Moreover, particles can push  particles below them, potentially un-trapping these or affecting their direction of motion.  
To model these effects, we introduce scalar forces $f_{i\to j}$ on each outlet of occupied sites, which satisfy:
\be
\label{force}
f_{i\to j} = \max(f_{j'\to i}+\theta_{i\to j}+\theta,0)
\ee
where  $f_{j'\to i}$ is the force on the input bond $j'\to i$ along the same direction as $i\to j$, as depicted in Fig.~\ref{model}. Eq.(\ref{force}) captures that forces are positive for repulsive particles, and that particle $i$ exerts a larger force on toward site $j$ if the bond inclination $\theta_{i\to j}$ is large, or if other particles above $i$ are pushing  it in that direction. From our analysis below, we expect that the details of the interactions are not relevant, as long as the  direction of motion of one particle can depend on the presence of particles above it- an ingredient not present in \cite{Watson96, Watson97}.

% When there is no particle at site $j'$, $f_{j'\to i}=0$. 
%
%The particles move on the landscape according to the following dynamical rules:

We update the position of the particles as follows, see Fig.~\ref{model} for illustration. We first compute all the  forces  in the system.
Next we consider one row of $W$ sites, and consider the motion of its  particles. Priority is set by considering  first outlets with the largest $f_{i\to j}$ and unoccupied downhill site $j$.   Once all possible moves ( $f_{i\to j}>0$, $j$ empty) have been made, forces are computed again in the system, 
and the next uphill row of particles is updated. When the $L$ rows forming the periodic system have all been updated, time increases by one.

For  given parameters $\theta,n$ we prepare the system via  two protocols.  In the ``quenched'' protocol, one considers a given frozen background, and launch the numerics with a large $\theta$ and randomly placed particles - parameters are such that the system  is well within the flowing phase. Next, $\theta$ is lowered slowly so that stationarity is always achieved. We also consider the ``Equilibrated'' protocol: for {\it any} $\theta$,  particles initial positions are  random. Dynamical properties are measured after the memory of the random initial condition is lost. We find that using different protocols does not change critical exponents, but that the quenched protocol appears to converge more slowly with system size. % $\theta_c$. % In what follows we make use of the ``quenched'' protocol for  explaining some of the results. 
Below we present most of our results obtained from the ``equilibrated'' protocol with $W=4\sqrt{L}$~\cite{Watson97}, and $n=0.25$ unless specified.

\begin{figure}[!ht]
\centering
\includegraphics[width=1.0\columnwidth]{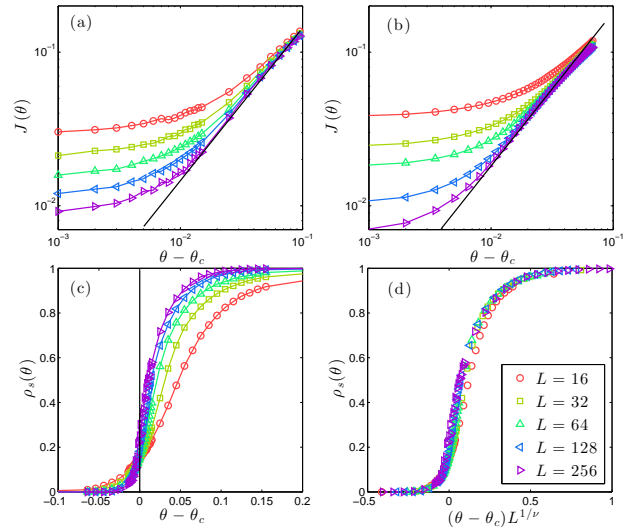}
\caption{\small{%Properties of the eroding phase of the model for different system sizes.
Average current $J$ versus $\theta-\theta_c$ in log-log scale for the (a) ``equilibrated'' and (b) ``quenched'' protocols, for which $\theta_c=0.164\pm0.002$ and $\theta_c=0.172\pm0.002$ respectively-  a difference plausibly due to finite size effects. The black solid lines with slope one indicate the linear relation  $J\propto \theta-\theta_c$. (c) Density of conducting sites  $\rho_s$ versus $\theta-\theta_c$ for the ``equilibrated'' protocol. (d) $\rho_s$ curves collapsed by rescaling $\theta-\theta_c$ with $L^{1/\nu}$, where $\nu=3.0\pm0.2$. %(e) Correlation in transversal direction $C_T$ of the conduction $\sigma$ of the sites in the steady state, the transversal distance $x$ is rescaled by the system with $W$.
}}\label{J}
\end{figure}

{\it Results:} Once the steady state is reached, we measure the average current of particles $J$ and the number density of sites carrying a finite current $\rho_s$.  Measurements of both quantities indicate a  sharp dynamical transition at some $\theta_c$ below which $J=0$ and $\rho_s=0$ as $L\rightarrow \infty$, see Fig.~\ref{model}. $\theta_c$ can be accurately extracted by considering the crossing point of the curves $\rho_s(\theta)$ as $L$ is varied, yielding $\theta_c=0.164\pm0.002$  for the equilibrated protocol. In the limit $L\rightarrow \infty$ our data extrapolates to:
\ba
\label{Jthe}
J(\theta)&\sim& \theta-\theta_c \ \  \hbox{ for} \ \theta>\theta_c\\
\label{rhot}
\rho_s(\theta)&=&\Theta(\theta-\theta_c),
\ea
where $\Theta$ is the Heaviside function. Eq.(\ref{Jthe}) corresponds to $\beta=1$, whereas Eq.(\ref{rhot}) indicates that all sites are visited by particles in the flowing phase.
Introducing the exponent $\rho_s(\theta)\sim (\theta-\theta_c)^\gamma$, this corresponds to $\gamma=0$.  The collapse of Fig.~\ref{J}(d) shows how convergence to Eq.(\ref{rhot}) takes place as $L\rightarrow \infty$, from which a finite size scaling length $\xi \sim (\theta-\theta_c)^{-\nu}$ with $\nu\approx 3$ can be extracted.

\begin{figure}[!ht]
\centering
\includegraphics[width=0.7\columnwidth]{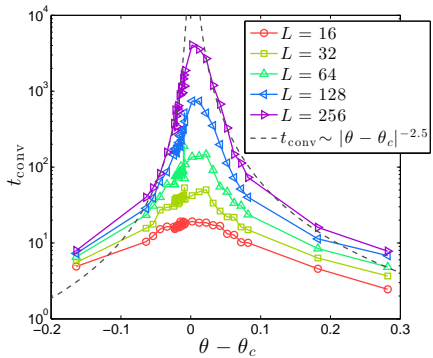}
\caption{\small{
Transient time $t_{\rm conv}$ {\it v.s.}  $\theta$. For a given realization, $t_{\rm conv}$ is defined as the smallest time for which $J(t)-J \leq \sqrt{Var(J)}$ where $Var(J)=\lim_{T\to\infty}\frac{1}{T}\sum_{t=1}^{T}(J(t)-J)^2$. The gray dashed lines correspond to $t_{\rm conv}\sim|\theta-\theta_c|^{-2.5}$.
}}\label{tau}
\end{figure}

Criticality is also observed in the transient time $t_{\rm conv}$ needed for the current to reach its stationary value.
Fig.~\ref{tau} reports that $t_{\rm conv}\sim |\theta-\theta_c|^{-z}$ with $z\approx 2.5$ on both sides of the transition. A similar exponent was observed numerically in \cite{Clark15}. %We also find that the dynamics becomes more intermittent as $\theta\rightarrow \theta_c$, as indicated by the rescaled variance of the flux in stationary state shown in Fig.\ref{tau}.a.

\begin{figure}[!ht]
\centering
\includegraphics[width=1.0\columnwidth]{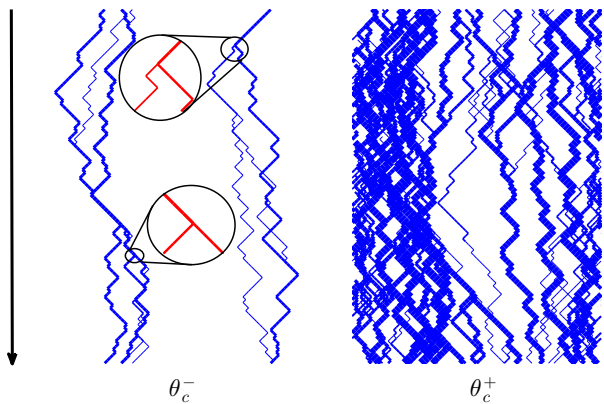}
\caption{\small{ Examples of drainage pattern  just below $\theta_c$ (Left) and above (Right). The black arrow shows the downhill direction. The thickness of the lines represents $\sigma_{i\to j}$ in logarithmic scale. A few examples showing  splitting events are magnified on the left. Here $W=45$ and $L=128$, and $J>0$ even below  $\theta_c$ due to finite size effects.}}\label{drain}
\end{figure}

The spatial organization of the current in steady state can be studied by considering the time-averaged local current $\sigma_i$ on site $i$, or the time-averaged  outlet current $\sigma_{i\to j}$. %(\textcolor{red}{$\sigma_i$ is the sum of the two outlets.}) 
The spatial average of each quantity is $J$.  Fig.~\ref{drain}  shows  an example of drainage pattern, i.e.  one realization of the map of the $\sigma_{i\to j}$.

%as the number of eroding particles passing through the site in a unit time, which is a static quantity in steady state for a given landscape and a certain drive $\theta$. Typical drainage patterns below and above threshold $\theta_c$ are shown in Fig.~\ref{drain}.The mean of the conductions is the current.

\begin{figure}[!ht]
\centering
\includegraphics[width=1.0\columnwidth]{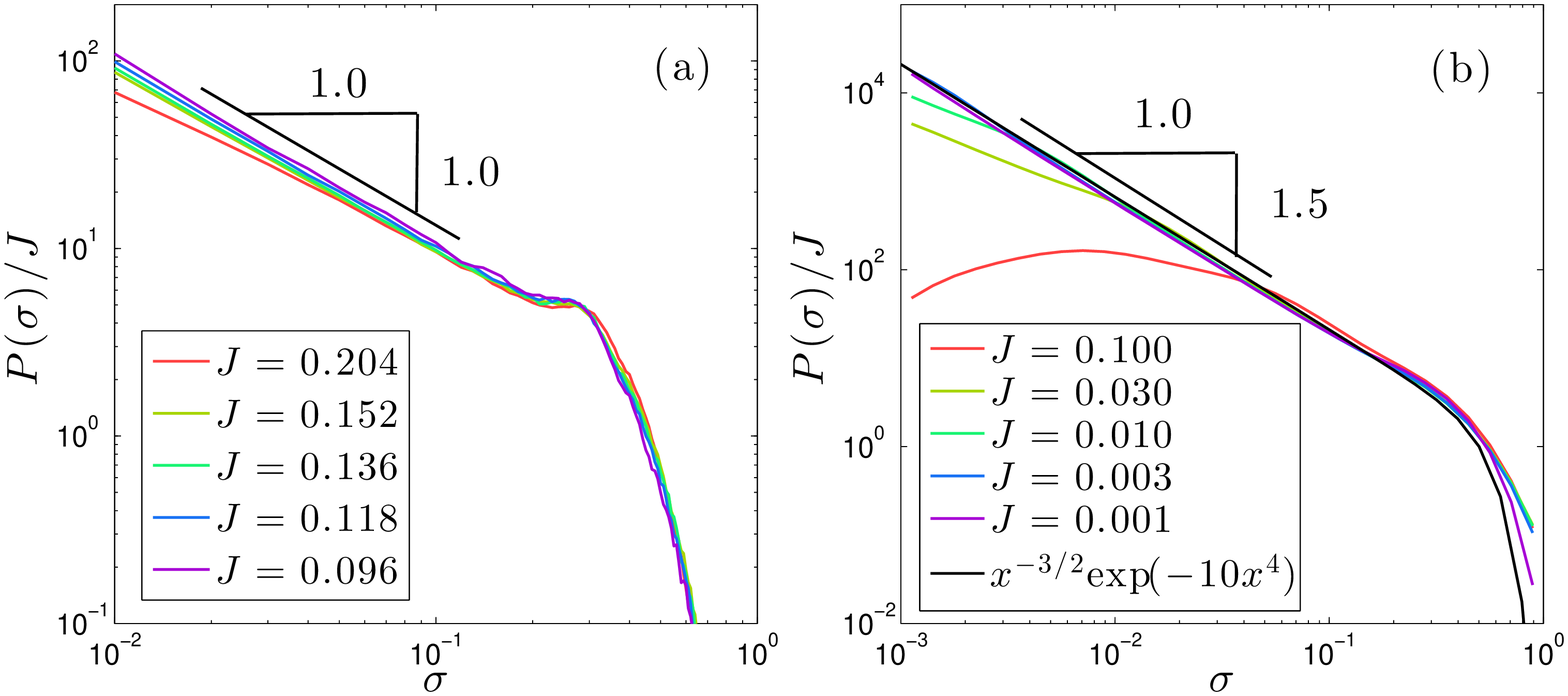}
\caption{\small{Distribution of the site current $P(\sigma)$ in steady state for given average currents $J$ of (a) the erosion model ($L=256$, $W=64$) and (b) our mean-field model ($W=1600$).}}\label{dist}
\end{figure}

To quantify such  patterns, we compute in Fig.~\ref{dist}(a) the distribution $P(\sigma)$ of the local current $\sigma_i$ for various mean current $J$.
We observed that:
\be
\label{psig}
P(\sigma)= J\sigma^{-\tau}f(\sigma)
\ee
where $\tau\approx1$ and $f$ is a cut-off function, expected since in our model $\sigma_i<1$. Eq.(\ref{psig}) indicates that $P(\sigma)$  is remarkably broad.
In fact, the divergence at small $\sigma$ is so pronounced that a cut-off $\sigma_{\min}$ must be present in Eq.(\ref{psig}) to guarantee a proper normalization of the distribution $P(\sigma)$, although we cannot detect it numerically. 

%, and expect it to be of order $\sigma_{\rm min}\sim J^{1/(\tau-1)}$ and even exponentially small if $\tau=1$.

%Remarkably, we find that  $P(\sigma)$ is extremely broad: it is power-law distributed, with a cut-off for $\sigma_i\sim 1$ The conduction is broadly distributed with a weakly diverging tail at small conduction, and sharply cut off at conduction is of order one. Remarkably, the distribution at different current can be perfectly collapsed by rescaling the distribution by the average current $J$, as shown in Fig.~\ref{dist}(a). We propose the following function for the distribution,
%
%where $\sigma_{\rm max}\sim1$ as particles are hard, and $\sigma_{\rm min}$ sets the lower cutoff when $\tau\geq1$ to ensure the distribution is normalizable, $\sigma_{\rm min}\sim J^{1/(\tau-1)}$. This form indicates that when $\tau<1$, $\rho_s\to0$ as $J\to0$ at $\theta_c$. The numerical results and the ubiquitous erosion strongly suggest that $\tau=1$.
% is the most relevant property. 
%It is straightforward to see that  both $J$ and $\rho_s$ are just the first and zeroth moments of the distribution,
%\begin{subequations}
%\begin{align}
%J = \int_{0^+}^{\infty}\sigma P(\sigma)\rd\sigma,
%\label{moma}\\
%\rho_s=\int_{0^+}^{\infty} P(\sigma)\rd\sigma. \label{momb}
%\end{align}
%\end{subequations}

\begin{figure}[!ht]
\centering
\includegraphics[width=1.0\columnwidth]{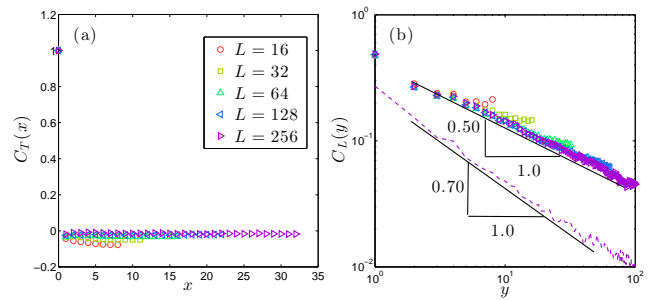}
\caption{\small{(a) Transverse current correlations $C_T$ at $\theta_c$  and (b) longitudinal current correlation   $C_L$ at $\theta_c$ and at $\theta-\theta_c=0.25$ for $L=256$ (dashed line).}}\label{core}
\end{figure}

Next, we compute the spatial correlation of the local current  in the transverse direction $C_T(x)$, defined as:
\be
\label{corr}
C_T(x)=\overline{(\langle\sigma_i\sigma_{i+x}\rangle-J^2)/(\langle\sigma_i^2\rangle-J^2)}
\ee 
where the site $i$ and $i+x$ are on the same row, but at a distance $x$ of each other. Here the brackets denote the spatial average, whereas the overline indicates averaging over the quenched randomness (the $h_i$'s).  Fig.~\ref{core}(a) shows that no transverse correlations exist for distances larger that  one site. However, long-range, power-law correlations are observed in the longitudinal direction, as can be seen by defining a longitudinal correlation function  $C_L(y)$, where $y$ is the vertical distance between two sites belonging to the same column. We find that $C_L(y)\sim 1/\sqrt{y}$ at $\theta_c$, but that $C_L(y)$ decays  somewhat faster deeper in the flowing phase, as shown in Fig.~\ref{core}(b).

%Next, we compute the spatial correlation of the local current  in the transverse direction $C_{T\textcolor{red}{,L}}(x\textcolor{red}{,y})$, defined as:
%\be
%\label{corr}
%C_{T\textcolor{red}{,L}}(x\textcolor{red}{,y})=\overline{(\langle\sigma_i\sigma_{i+x\textcolor{red}{,y}}\rangle-J^2)/(\langle\sigma_i^2\rangle-J^2)}
%\ee 
%where the site $i$ and $i+x$ are on the same row, but at a distance $x$ of each other, \textcolor{red}{and $y$ is the vertical distance between two sites belonging to the same column.} Here the brackets denote the spatial average, whereas the overline indicates averaging over the quenched randomness (the $h_i$'s).  Fig.~\ref{core}(a) shows that no transverse correlations exist for distances larger that  one site. However, long-range, power-law correlations are observed in the longitudinal direction, \textcolor{red}{and the exponent can vary for different $\theta$ in flowing phase.}%, as can be seen by defining a longitudinal correlation function  $C_L(y)$, where . 

{\it A scaling relation:} we now derive a relationship between the exponents $\beta$ characterizing $J$ and $\gamma$  characterizing $\rho_s$. It holds true for both protocols, but is presented here in 
the ``quenched" case. Near threshold, at any instant of time the density of moving particles is $J\ll n<1$, thus most of the particles are trapped and will move only when a mobile particle passes by. As $\theta$ is decreased by some $\delta\theta$, a finite density of new traps $\delta m\sim \delta \theta$ is created. If these traps appear on the region of size $\rho_s$ where mobile particles flow, they will reduce the fraction of mobile particle by $\delta J= \rho_s \delta m\sim \rho_s \delta\theta$, which implies:
\be
\label{screlation}
\beta=\gamma+1
\ee
%For the ``equilibrated'' protocol, particles need to fill all traps in the drainage basin~\cite{Narayan94,Watson97}, $\phi$, the landscape explored by mobile particles in transient. $\phi\geq\rho_s=1$ when $J>0$, and $\phi=0$ if $J=0$.  
Eq.(\ref{screlation}) shows that the result $\beta=1$ is a direct consequence of the fact that in our model, all sites are explored by mobile particles for $\theta>\theta_c$, a result which is not obvious. In the dirty superconductor models of  \cite{Narayan94,Watson97}, this is not the case and for the ``equilibrated" protocol  $\beta>1$ was found. We argue that this difference comes from the dynamical rules chosen in \cite{Narayan94,Watson97}, according to which ``rivers" forming the drainage pattern never split: their current grows in amplitude in the downhill direction, until it reaches unity. In these models the drainage pattern thus consists of rivers of unit current, avoiding each other, and separated by a typical distance of order $1/J$. Our model behaves completely differently because rivers can split, as emphasized in Fig.~\ref{drain}. This comes about because the direction taken by a particle can depend on the presence of a particle right above it, as illustrated in case A of Fig.~\ref{model}. This effect is expected to occur in the erosion problem due to hydrodynamic interactions or direct contact between particles, and may also be relevant for superconductors.

{\it Mean-field model:} we now seek to quantify the effect of splitting. Its relevance is not obvious a priori, as splitting stems from particle interactions, 
and may thus become less important as the fraction of moving particles vanishes as $J\rightarrow 0$. To model this effect we consider that the current $\sigma_i$  on a site $i$ is decomposed in its two outlets as $\sigma_i=q \sigma_i +(1-q)\sigma_i$, where $q$ is a random variable of distribution $\eta(q)$. If there were no splitting then $\eta(q)=\frac{1}{2}\delta(q)+\frac{1}{2}\delta(1-q)$. Here instead, we assume that $\eta(q)=\frac{1}{2}\delta(q-J)+\frac{1}{2}\delta(1-J-q)$. %(\textcolor{red}{I tried to explain why the split is proportional to $J$ in my initial version with both case A and B. Do you think it's too complicated with little information?}) 
This choice captures that the probability of splitting is increased if more moving particles are present, and can occur for example if two particles flow behind each other, as exemplified in case A of Fig.~\ref{model}. Next, we make the mean field assumption that two adjacent sites $i$ and $j$ on the same row are uncorrelated, $P(\sigma_i,\sigma_j)=P(\sigma_i)P(\sigma_j)$. We then obtain the self-consistent equation  that $P(\sigma)$ must be equal to:
\be
\label{selfconsist}
%P(\sigma)=
\int\rd q_1\rd q_2\rd\sigma_1\rd\sigma_2\eta(q_1)\eta(q_2)
P(\sigma_1)P(\sigma_2)\delta(q_1\sigma_1+q_2\sigma_2-q) 
\ee
This mean-field model belongs to the class of $q$-models introduced to study force propagation \cite{Liu95, Coppersmith96}. It is easy to simulate, and some aspects of the solution can be computed. Numerical results are shown in Fig.~\ref{dist}(b). The result obtained for $P(\sigma)$ is very similar to Eq.(\ref{psig}) that describes our erosion model: $P(\sigma)$ is found to be power-law distributed (although $\tau=3/2$ instead of $\tau=1$) where with an upper cutoff at $\sigma_{\max}\sim 1$, and $P(\sigma) \propto J$. 

These results are of interest, as they explain why $P(\sigma)$ is very broad,  and is not dominated by sites displaying no current at all (which would correspond to a delta function at zero) even as $J\rightarrow 0$, thus confirming that $\gamma=0$. They can be explained by taking the Laplace transform $\tilde P$ of Eq.(\ref{selfconsist}). One then obtains a non-linear differential equation for $\tilde P$, from which it can be argued generically that $\tau=3/2$ \cite{Coppersmith96}. We have performed a Taylor expansion of $\tilde P$ around zero, which leads to relationship between the different moments of the distribution $P(\sigma)$. From it,
we can show that $P(\sigma)\propto J$ and $\sigma_{\max}\sim 1$. We also find that the cut-off  of the divergence of $P(\sigma)$ at small argument follows $\sigma_{\min}\sim J^{1/(\tau-1)}$.

{\it Conclusion:} we have introduced a novel model for over-damped interacting particles driven on a disordered substrate.
It predicts a dynamical phase transition at some threshold forcing $\theta_c$, and makes quantitative predictions  for various quantities including  the particle current  and the drainage pattern.  The latter  could be tested experimentally in erosion experiments \cite{Charru04,Ouriemi09,Houssais15, Lobkovsky08} by tracking particles on the surface \cite{Charru04} to reconstruct the spatial organization of current. Another interesting set-up  are colloids at an interface, pinned by a random environment generated by a rough charged surface \cite{Pertsinidis08}. Numerics support the existence of a dynamical transition in this system  where flow localizes on  channels \cite{Reichhardt02}, which may fall in the universality class of our model.
%Our prediction that the value of $\theta_c$ depends on the protocol, and is large if $\theta$ is decreased continuously, would also be interesting to test.

Note that our model assumes that particles are over-damped, and that their inertia is negligible. 
We expect inertia to lead to hysteresis and make the transition first order, as observed on inertial granular flows down an inclined plane \cite{Andreotti13}, although this effect may be small in practice, as supported by experiments \cite{Ouriemi07}. We did not consider non-laminar flows, nor temperature (that can be relevant for colloids). Both effects should smooth  the transition, and lead to creep even below $\theta_c$.% \cite{Houssais15}.

Finally, it has been proposed that the erosion threshold is a dynamical transition very similar to the jamming transition that occurs when a bulk amorphous material is sheared \cite{Houssais15}. If our model holds, this is not the case: due to the presence of the free interface, long-range elastic interactions between mobile particles are absent. In recent theoretical descriptions of the jamming transition such  interactions are central   both for soft \cite{Lin2014} and hard particles \cite{Lerner12a,DeGiuli14d}.

\begin{acknowledgements}
We thank B.~ Andreotti, P.~Aussilous, M.~Baity-Jesi, D. Bartolo, E.~DeGiuli, E.~Guazzelli, J.~Lin, B. ~Metzger and Y.~Rabin for discussions. This work has been supported primarily by the National Science Foundation CBET-1236378 and MRSEC Program of the NSF DMR-0820341 for partial
funding.
\end{acknowledgements}

\bibliography{Wyartbibnew}

\begin{thebibliography}{25}%
\makeatletter
\providecommand \@ifxundefined [1]{%
 \@ifx{#1\undefined}
}%
\providecommand \@ifnum [1]{%
 \ifnum #1\expandafter \@firstoftwo
 \else \expandafter \@secondoftwo
 \fi
}%
\providecommand \@ifx [1]{%
 \ifx #1\expandafter \@firstoftwo
 \else \expandafter \@secondoftwo
 \fi
}%
\providecommand \natexlab [1]{#1}%
\providecommand \enquote  [1]{``#1''}%
\providecommand \bibnamefont  [1]{#1}%
\providecommand \bibfnamefont [1]{#1}%
\providecommand \citenamefont [1]{#1}%
\providecommand \href@noop [0]{\@secondoftwo}%
\providecommand \href [0]{\begingroup \@sanitize@url \@href}%
\providecommand \@href[1]{\@@startlink{#1}\@@href}%
\providecommand \@@href[1]{\endgroup#1\@@endlink}%
\providecommand \@sanitize@url [0]{\catcode `\\12\catcode `\$12\catcode
  `\&12\catcode `\#12\catcode `\^12\catcode `\_12\catcode `\%12\relax}%
\providecommand \@@startlink[1]{}%
\providecommand \@@endlink[0]{}%
\providecommand \url  [0]{\begingroup\@sanitize@url \@url }%
\providecommand \@url [1]{\endgroup\@href {#1}{\urlprefix }}%
\providecommand \urlprefix  [0]{URL }%
\providecommand \Eprint [0]{\href }%
\providecommand \doibase [0]{http://dx.doi.org/}%
\providecommand \selectlanguage [0]{\@gobble}%
\providecommand \bibinfo  [0]{\@secondoftwo}%
\providecommand \bibfield  [0]{\@secondoftwo}%
\providecommand \translation [1]{[#1]}%
\providecommand \BibitemOpen [0]{}%
\providecommand \bibitemStop [0]{}%
\providecommand \bibitemNoStop [0]{.\EOS\space}%
\providecommand \EOS [0]{\spacefactor3000\relax}%
\providecommand \BibitemShut  [1]{\csname bibitem#1\endcsname}%
\let\auto@bib@innerbib\@empty
%</preamble>
\bibitem [{\citenamefont {Shields}(1936)}]{Shields36}%
  \BibitemOpen
  \bibfield  {author} {\bibinfo {author} {\bibfnamefont {A.}~\bibnamefont
  {Shields}},\ }\href@noop {} {\bibfield  {journal} {\bibinfo  {journal} {Mitt.
  Preuss. Vers. Anst. Wasserb. u. Schiffb., Berlin, Heft}\ ,\ \bibinfo {pages}
  {26}} (\bibinfo {year} {1936})}\BibitemShut {NoStop}%
\bibitem [{\citenamefont {White}(1970)}]{White70}%
  \BibitemOpen
  \bibfield  {author} {\bibinfo {author} {\bibfnamefont {S.~J.}\ \bibnamefont
  {White}},\ }\href {http://dx.doi.org/10.1038/228152a0} {\bibfield  {journal}
  {\bibinfo  {journal} {Nature}\ }\textbf {\bibinfo {volume} {228}},\ \bibinfo
  {pages} {152} (\bibinfo {year} {1970})}\BibitemShut {NoStop}%
\bibitem [{\citenamefont {Lobkovsky}\ \emph {et~al.}(2008)\citenamefont
  {Lobkovsky}, \citenamefont {Orpe}, \citenamefont {Molloy}, \citenamefont
  {Kudrolli},\ and\ \citenamefont {Rothman}}]{Lobkovsky08}%
  \BibitemOpen
  \bibfield  {author} {\bibinfo {author} {\bibfnamefont {A.~E.}\ \bibnamefont
  {Lobkovsky}}, \bibinfo {author} {\bibfnamefont {A.~V.}\ \bibnamefont {Orpe}},
  \bibinfo {author} {\bibfnamefont {R.}~\bibnamefont {Molloy}}, \bibinfo
  {author} {\bibfnamefont {A.}~\bibnamefont {Kudrolli}}, \ and\ \bibinfo
  {author} {\bibfnamefont {D.~H.}\ \bibnamefont {Rothman}},\ }\href {\doibase
  10.1017/S0022112008001389} {\bibfield  {journal} {\bibinfo  {journal}
  {Journal of Fluid Mechanics}\ }\textbf {\bibinfo {volume} {605}},\ \bibinfo
  {pages} {47} (\bibinfo {year} {2008})}\BibitemShut {NoStop}%
\bibitem [{\citenamefont {Charru}\ \emph {et~al.}(2004)\citenamefont {Charru},
  \citenamefont {Mouilleron},\ and\ \citenamefont {Eiff}}]{Charru04}%
  \BibitemOpen
  \bibfield  {author} {\bibinfo {author} {\bibfnamefont {F.}~\bibnamefont
  {Charru}}, \bibinfo {author} {\bibfnamefont {H.}~\bibnamefont {Mouilleron}},
  \ and\ \bibinfo {author} {\bibfnamefont {O.}~\bibnamefont {Eiff}},\ }\href
  {\doibase 10.1017/S0022112004001028} {\bibfield  {journal} {\bibinfo
  {journal} {Journal of Fluid Mechanics}\ }\textbf {\bibinfo {volume} {519}},\
  \bibinfo {pages} {55} (\bibinfo {year} {2004})}\BibitemShut {NoStop}%
\bibitem [{\citenamefont {Aussillous}\ \emph {et~al.}(2013)\citenamefont
  {Aussillous}, \citenamefont {Chauchat}, \citenamefont {Pailha}, \citenamefont
  {M{\'e}dale},\ and\ \citenamefont {Guazzelli}}]{Aussillous13}%
  \BibitemOpen
  \bibfield  {author} {\bibinfo {author} {\bibfnamefont {P.}~\bibnamefont
  {Aussillous}}, \bibinfo {author} {\bibfnamefont {J.}~\bibnamefont
  {Chauchat}}, \bibinfo {author} {\bibfnamefont {M.}~\bibnamefont {Pailha}},
  \bibinfo {author} {\bibfnamefont {M.}~\bibnamefont {M{\'e}dale}}, \ and\
  \bibinfo {author} {\bibfnamefont {{\'E}.}~\bibnamefont {Guazzelli}},\ }\href
  {\doibase 10.1017/jfm.2013.546} {\bibfield  {journal} {\bibinfo  {journal}
  {Journal of Fluid Mechanics}\ }\textbf {\bibinfo {volume} {736}},\ \bibinfo
  {pages} {594} (\bibinfo {year} {2013})}\BibitemShut {NoStop}%
\bibitem [{\citenamefont {Houssais}\ \emph {et~al.}(2015)\citenamefont
  {Houssais}, \citenamefont {Ortiz}, \citenamefont {Durian},\ and\
  \citenamefont {Jerolmack}}]{Houssais15}%
  \BibitemOpen
  \bibfield  {author} {\bibinfo {author} {\bibfnamefont {M.}~\bibnamefont
  {Houssais}}, \bibinfo {author} {\bibfnamefont {C.~P.}\ \bibnamefont {Ortiz}},
  \bibinfo {author} {\bibfnamefont {D.~J.}\ \bibnamefont {Durian}}, \ and\
  \bibinfo {author} {\bibfnamefont {D.~J.}\ \bibnamefont {Jerolmack}},\ }\href
  {http://dx.doi.org/10.1038/ncomms7527} {\bibfield  {journal} {\bibinfo
  {journal} {Nat Commun}\ }\textbf {\bibinfo {volume} {6}} (\bibinfo {year}
  {2015})}\BibitemShut {NoStop}%
\bibitem [{\citenamefont {Parker}\ \emph {et~al.}(2007)\citenamefont {Parker},
  \citenamefont {Wilcock}, \citenamefont {Paola}, \citenamefont {Dietrich},\
  and\ \citenamefont {Pitlick}}]{Parker07}%
  \BibitemOpen
  \bibfield  {author} {\bibinfo {author} {\bibfnamefont {G.}~\bibnamefont
  {Parker}}, \bibinfo {author} {\bibfnamefont {P.~R.}\ \bibnamefont {Wilcock}},
  \bibinfo {author} {\bibfnamefont {C.}~\bibnamefont {Paola}}, \bibinfo
  {author} {\bibfnamefont {W.~E.}\ \bibnamefont {Dietrich}}, \ and\ \bibinfo
  {author} {\bibfnamefont {J.}~\bibnamefont {Pitlick}},\ }\href {\doibase
  10.1029/2006JF000549} {\bibfield  {journal} {\bibinfo  {journal} {Journal of
  Geophysical Research: Earth Surface}\ }\textbf {\bibinfo {volume} {112}},\
  \bibinfo {pages} {n/a} (\bibinfo {year} {2007})}\BibitemShut {NoStop}%
\bibitem [{\citenamefont {Bagnold}(1966)}]{Bagnold66}%
  \BibitemOpen
  \bibfield  {author} {\bibinfo {author} {\bibfnamefont {R.~A.}\ \bibnamefont
  {Bagnold}},\ }\href@noop {} {\emph {\bibinfo {title} {The Physics of Sediment
  Transport by Wind and Water: A Collection of Hallmark Papers by RA
  Bagnold}}},\ \bibinfo {number} {231}\ (\bibinfo {year} {1966})\BibitemShut
  {NoStop}%
\bibitem [{\citenamefont {Ouriemi}\ \emph {et~al.}(2009)\citenamefont
  {Ouriemi}, \citenamefont {Aussillous},\ and\ \citenamefont
  {Guazzelli}}]{Ouriemi09}%
  \BibitemOpen
  \bibfield  {author} {\bibinfo {author} {\bibfnamefont {M.}~\bibnamefont
  {Ouriemi}}, \bibinfo {author} {\bibfnamefont {P.}~\bibnamefont {Aussillous}},
  \ and\ \bibinfo {author} {\bibfnamefont {E.}~\bibnamefont {Guazzelli}},\
  }\href {\doibase 10.1017/S0022112009007915} {\bibfield  {journal} {\bibinfo
  {journal} {Journal of Fluid Mechanics}\ }\textbf {\bibinfo {volume} {636}},\
  \bibinfo {pages} {295} (\bibinfo {year} {2009})}\BibitemShut {NoStop}%
\bibitem [{\citenamefont {Lajeunesse}\ \emph {et~al.}(2010)\citenamefont
  {Lajeunesse}, \citenamefont {Malverti},\ and\ \citenamefont
  {Charru}}]{Lajeunesse10}%
  \BibitemOpen
  \bibfield  {author} {\bibinfo {author} {\bibfnamefont {E.}~\bibnamefont
  {Lajeunesse}}, \bibinfo {author} {\bibfnamefont {L.}~\bibnamefont
  {Malverti}}, \ and\ \bibinfo {author} {\bibfnamefont {F.}~\bibnamefont
  {Charru}},\ }\href@noop {} {\bibfield  {journal} {\bibinfo  {journal}
  {Journal of Geophysical Research: Earth Surface (2003--2012)}\ }\textbf
  {\bibinfo {volume} {115}} (\bibinfo {year} {2010})}\BibitemShut {NoStop}%
\bibitem [{\citenamefont {Leighton}\ and\ \citenamefont
  {Acrivos}(1986)}]{Leighton86}%
  \BibitemOpen
  \bibfield  {author} {\bibinfo {author} {\bibfnamefont {D.}~\bibnamefont
  {Leighton}}\ and\ \bibinfo {author} {\bibfnamefont {A.}~\bibnamefont
  {Acrivos}},\ }\href {\doibase http://dx.doi.org/10.1016/0009-2509(86)85225-3}
  {\bibfield  {journal} {\bibinfo  {journal} {Chemical Engineering Science}\
  }\textbf {\bibinfo {volume} {41}},\ \bibinfo {pages} {1377 } (\bibinfo {year}
  {1986})}\BibitemShut {NoStop}%
\bibitem [{\citenamefont {Watson}\ and\ \citenamefont
  {Fisher}(1996)}]{Watson96}%
  \BibitemOpen
  \bibfield  {author} {\bibinfo {author} {\bibfnamefont {J.}~\bibnamefont
  {Watson}}\ and\ \bibinfo {author} {\bibfnamefont {D.~S.}\ \bibnamefont
  {Fisher}},\ }\href {\doibase 10.1103/PhysRevB.54.938} {\bibfield  {journal}
  {\bibinfo  {journal} {Phys. Rev. B}\ }\textbf {\bibinfo {volume} {54}},\
  \bibinfo {pages} {938} (\bibinfo {year} {1996})}\BibitemShut {NoStop}%
\bibitem [{\citenamefont {Watson}\ and\ \citenamefont
  {Fisher}(1997)}]{Watson97}%
  \BibitemOpen
  \bibfield  {author} {\bibinfo {author} {\bibfnamefont {J.}~\bibnamefont
  {Watson}}\ and\ \bibinfo {author} {\bibfnamefont {D.~S.}\ \bibnamefont
  {Fisher}},\ }\href {\doibase 10.1103/PhysRevB.55.14909} {\bibfield  {journal}
  {\bibinfo  {journal} {Phys. Rev. B}\ }\textbf {\bibinfo {volume} {55}},\
  \bibinfo {pages} {14909} (\bibinfo {year} {1997})}\BibitemShut {NoStop}%
\bibitem [{\citenamefont {Liu}\ \emph {et~al.}(1995)\citenamefont {Liu},
  \citenamefont {Nagel}, \citenamefont {Schecter}, \citenamefont {Coppersmith},
  \citenamefont {Majumdar}, \citenamefont {Narayan},\ and\ \citenamefont
  {Witten}}]{Liu95}%
  \BibitemOpen
  \bibfield  {author} {\bibinfo {author} {\bibfnamefont {C.~h.}\ \bibnamefont
  {Liu}}, \bibinfo {author} {\bibfnamefont {S.~R.}\ \bibnamefont {Nagel}},
  \bibinfo {author} {\bibfnamefont {D.~A.}\ \bibnamefont {Schecter}}, \bibinfo
  {author} {\bibfnamefont {S.~N.}\ \bibnamefont {Coppersmith}}, \bibinfo
  {author} {\bibfnamefont {S.}~\bibnamefont {Majumdar}}, \bibinfo {author}
  {\bibfnamefont {O.}~\bibnamefont {Narayan}}, \ and\ \bibinfo {author}
  {\bibfnamefont {T.~A.}\ \bibnamefont {Witten}},\ }\href {\doibase
  10.1126/science.269.5223.513} {\bibfield  {journal} {\bibinfo  {journal}
  {Science}\ }\textbf {\bibinfo {volume} {269}},\ \bibinfo {pages} {513}
  (\bibinfo {year} {1995})}\BibitemShut {NoStop}%
\bibitem [{\citenamefont {Coppersmith}\ \emph {et~al.}(1996)\citenamefont
  {Coppersmith}, \citenamefont {Liu}, \citenamefont {Majumdar}, \citenamefont
  {Narayan},\ and\ \citenamefont {Witten}}]{Coppersmith96}%
  \BibitemOpen
  \bibfield  {author} {\bibinfo {author} {\bibfnamefont {S.~N.}\ \bibnamefont
  {Coppersmith}}, \bibinfo {author} {\bibfnamefont {C.~h.}\ \bibnamefont
  {Liu}}, \bibinfo {author} {\bibfnamefont {S.}~\bibnamefont {Majumdar}},
  \bibinfo {author} {\bibfnamefont {O.}~\bibnamefont {Narayan}}, \ and\
  \bibinfo {author} {\bibfnamefont {T.~A.}\ \bibnamefont {Witten}},\ }\href
  {\doibase 10.1103/PhysRevE.53.4673} {\bibfield  {journal} {\bibinfo
  {journal} {Phys. Rev. E}\ }\textbf {\bibinfo {volume} {53}},\ \bibinfo
  {pages} {4673} (\bibinfo {year} {1996})}\BibitemShut {NoStop}%
\bibitem [{\citenamefont {Rinaldo}\ \emph {et~al.}(2014)\citenamefont
  {Rinaldo}, \citenamefont {Rigon}, \citenamefont {Banavar}, \citenamefont
  {Maritan},\ and\ \citenamefont {Rodriguez-Iturbe}}]{Rinaldo14}%
  \BibitemOpen
  \bibfield  {author} {\bibinfo {author} {\bibfnamefont {A.}~\bibnamefont
  {Rinaldo}}, \bibinfo {author} {\bibfnamefont {R.}~\bibnamefont {Rigon}},
  \bibinfo {author} {\bibfnamefont {J.~R.}\ \bibnamefont {Banavar}}, \bibinfo
  {author} {\bibfnamefont {A.}~\bibnamefont {Maritan}}, \ and\ \bibinfo
  {author} {\bibfnamefont {I.}~\bibnamefont {Rodriguez-Iturbe}},\ }\href
  {\doibase 10.1073/pnas.1322700111} {\bibfield  {journal} {\bibinfo  {journal}
  {Proceedings of the National Academy of Sciences}\ }\textbf {\bibinfo
  {volume} {111}},\ \bibinfo {pages} {2417} (\bibinfo {year}
  {2014})}\BibitemShut {NoStop}%
\bibitem [{\citenamefont {Clark}\ \emph {et~al.}(2015)\citenamefont {Clark},
  \citenamefont {Shattuck}, \citenamefont {Ouellette},\ and\ \citenamefont
  {O'Hern}}]{Clark15}%
  \BibitemOpen
  \bibfield  {author} {\bibinfo {author} {\bibfnamefont {A.~H.}\ \bibnamefont
  {Clark}}, \bibinfo {author} {\bibfnamefont {M.~D.}\ \bibnamefont {Shattuck}},
  \bibinfo {author} {\bibfnamefont {N.~T.}\ \bibnamefont {Ouellette}}, \ and\
  \bibinfo {author} {\bibfnamefont {C.~S.}\ \bibnamefont {O'Hern}},\
  }\href@noop {} {\bibfield  {journal} {\bibinfo  {journal} {arXiv preprint
  arXiv:1504.03403}\ } (\bibinfo {year} {2015})}\BibitemShut {NoStop}%
\bibitem [{\citenamefont {Narayan}\ and\ \citenamefont
  {Fisher}(1994)}]{Narayan94}%
  \BibitemOpen
  \bibfield  {author} {\bibinfo {author} {\bibfnamefont {O.}~\bibnamefont
  {Narayan}}\ and\ \bibinfo {author} {\bibfnamefont {D.~S.}\ \bibnamefont
  {Fisher}},\ }\href {\doibase 10.1103/PhysRevB.49.9469} {\bibfield  {journal}
  {\bibinfo  {journal} {Phys. Rev. B}\ }\textbf {\bibinfo {volume} {49}},\
  \bibinfo {pages} {9469} (\bibinfo {year} {1994})}\BibitemShut {NoStop}%
\bibitem [{\citenamefont {Pertsinidis}\ and\ \citenamefont
  {Ling}(2008)}]{Pertsinidis08}%
  \BibitemOpen
  \bibfield  {author} {\bibinfo {author} {\bibfnamefont {A.}~\bibnamefont
  {Pertsinidis}}\ and\ \bibinfo {author} {\bibfnamefont {X.~S.}\ \bibnamefont
  {Ling}},\ }\href@noop {} {\bibfield  {journal} {\bibinfo  {journal} {Physical
  review letters}\ }\textbf {\bibinfo {volume} {100}},\ \bibinfo {pages}
  {028303} (\bibinfo {year} {2008})}\BibitemShut {NoStop}%
\bibitem [{\citenamefont {Reichhardt}\ and\ \citenamefont
  {Olson}(2002)}]{Reichhardt02}%
  \BibitemOpen
  \bibfield  {author} {\bibinfo {author} {\bibfnamefont {C.}~\bibnamefont
  {Reichhardt}}\ and\ \bibinfo {author} {\bibfnamefont {C.}~\bibnamefont
  {Olson}},\ }\href@noop {} {\bibfield  {journal} {\bibinfo  {journal}
  {Physical review letters}\ }\textbf {\bibinfo {volume} {89}},\ \bibinfo
  {pages} {078301} (\bibinfo {year} {2002})}\BibitemShut {NoStop}%
\bibitem [{\citenamefont {Andreotti}\ \emph {et~al.}(2013)\citenamefont
  {Andreotti}, \citenamefont {Forterre},\ and\ \citenamefont
  {Pouliquen}}]{Andreotti13}%
  \BibitemOpen
  \bibfield  {author} {\bibinfo {author} {\bibfnamefont {B.}~\bibnamefont
  {Andreotti}}, \bibinfo {author} {\bibfnamefont {Y.}~\bibnamefont {Forterre}},
  \ and\ \bibinfo {author} {\bibfnamefont {O.}~\bibnamefont {Pouliquen}},\
  }\href@noop {} {\emph {\bibinfo {title} {Granular media: between fluid and
  solid}}}\ (\bibinfo  {publisher} {Cambridge University Press},\ \bibinfo
  {year} {2013})\BibitemShut {NoStop}%
\bibitem [{\citenamefont {Ouriemi}\ \emph {et~al.}(2007)\citenamefont
  {Ouriemi}, \citenamefont {Aussillous}, \citenamefont {Medale}, \citenamefont
  {Peysson},\ and\ \citenamefont {Guazzelli}}]{Ouriemi07}%
  \BibitemOpen
  \bibfield  {author} {\bibinfo {author} {\bibfnamefont {M.}~\bibnamefont
  {Ouriemi}}, \bibinfo {author} {\bibfnamefont {P.}~\bibnamefont {Aussillous}},
  \bibinfo {author} {\bibfnamefont {M.}~\bibnamefont {Medale}}, \bibinfo
  {author} {\bibfnamefont {Y.}~\bibnamefont {Peysson}}, \ and\ \bibinfo
  {author} {\bibfnamefont {{\'E}.}~\bibnamefont {Guazzelli}},\ }\href@noop {}
  {\bibfield  {journal} {\bibinfo  {journal} {Physics of Fluids}\ }\textbf
  {\bibinfo {volume} {19}},\ \bibinfo {pages} {61706} (\bibinfo {year}
  {2007})}\BibitemShut {NoStop}%
\bibitem [{\citenamefont {Lin}\ \emph {et~al.}(2014)\citenamefont {Lin},
  \citenamefont {Lerner}, \citenamefont {Rosso},\ and\ \citenamefont
  {Wyart}}]{Lin2014}%
  \BibitemOpen
  \bibfield  {author} {\bibinfo {author} {\bibfnamefont {J.}~\bibnamefont
  {Lin}}, \bibinfo {author} {\bibfnamefont {E.}~\bibnamefont {Lerner}},
  \bibinfo {author} {\bibfnamefont {A.}~\bibnamefont {Rosso}}, \ and\ \bibinfo
  {author} {\bibfnamefont {M.}~\bibnamefont {Wyart}},\ }\href@noop {}
  {\bibfield  {journal} {\bibinfo  {journal} {Proceedings of the National
  Academy of Sciences}\ }\textbf {\bibinfo {volume} {111}},\ \bibinfo {pages}
  {14382} (\bibinfo {year} {2014})}\BibitemShut {NoStop}%
\bibitem [{\citenamefont {Lerner}\ \emph {et~al.}(2012)\citenamefont {Lerner},
  \citenamefont {D\"uring},\ and\ \citenamefont {Wyart}}]{Lerner12a}%
  \BibitemOpen
  \bibfield  {author} {\bibinfo {author} {\bibfnamefont {E.}~\bibnamefont
  {Lerner}}, \bibinfo {author} {\bibfnamefont {G.}~\bibnamefont {D\"uring}}, \
  and\ \bibinfo {author} {\bibfnamefont {M.}~\bibnamefont {Wyart}},\
  }\href@noop {} {\bibfield  {journal} {\bibinfo  {journal} {Proceedings of the
  National Academy of Sciences}\ }\textbf {\bibinfo {volume} {109}},\ \bibinfo
  {pages} {4798} (\bibinfo {year} {2012})}\BibitemShut {NoStop}%
\bibitem [{\citenamefont {DeGiuli}\ \emph {et~al.}(2014)\citenamefont
  {DeGiuli}, \citenamefont {D{\"u}ring}, \citenamefont {Lerner},\ and\
  \citenamefont {Wyart}}]{DeGiuli14d}%
  \BibitemOpen
  \bibfield  {author} {\bibinfo {author} {\bibfnamefont {E.}~\bibnamefont
  {DeGiuli}}, \bibinfo {author} {\bibfnamefont {G.}~\bibnamefont {D{\"u}ring}},
  \bibinfo {author} {\bibfnamefont {E.}~\bibnamefont {Lerner}}, \ and\ \bibinfo
  {author} {\bibfnamefont {M.}~\bibnamefont {Wyart}},\ }\href@noop {}
  {\bibfield  {journal} {\bibinfo  {journal} {arXiv preprint arXiv:1410.3535}\
  } (\bibinfo {year} {2014})}\BibitemShut {NoStop}%
\end{thebibliography}%

\end{document}